# Raman spectra of unfilled and filled carbon nanotubes: Theory

S. Gayen[1], S.N. Behera[2] and S. M. Bose[1+]

[1]Department of Physics, Drexel University, Philadelphia, PA 19104, USA

[2]Institute of Materials Science, Planetarium Building, Bhubaneswar-751013, Orissa, India.

## ABSTRACT

The Raman spectra of two G-bands and a radial breathing mode (RBM) of unfilled and filled single-wall semiconducting and metallic carbon nanotubes have been investigated theoretically, in the presence of electron-phonon and phonon-phonon interactions. Excitation of low frequency optical plasmons in the metallic nanotube is responsible for the peak known as the Breit-Wigner-Fano (BWF) line shape in the G-band Raman spectra. In a filled nanotube there is an additional peak due to excitation of the phonon of the filling atom or molecule. Positions, shapes and relative strengths of these Raman peaks depend on the phonon frequencies of the nanotube and that of the filling atoms, and strengths and forms of the plasmon-phonon and phonon-phonon interactions. For example, filling atoms with phonon frequency close to the RBM frequency of the nanotube may broaden and lower the RBM Raman peak to such an extent that it may become barely visible. Hybridization between the G-bands and the filling atom phonon is also strong when these two frequencies are close to each other and it has important effects on the G-band and the BWF line shapes. When the phonon frequency of the filling atom is far from the RBM and G-band frequencies, it gives rise to a separate peak with modest effects on the RBM and G-band spectra. Raman spectra of semiconducting unfilled and filled nanotubes have similar behavior as those of metallic nanotubes except that normally they have Lorentzian line shapes and do not show a BWF line shape. However, if a semiconducting nanotube is filled with donor atoms, it is predicted that the BWF type line shape may be observed near the RBM, or the G-band or the filling atom Raman peak.

# I. INTRODUCTION

Ever since the discovery of carbon nanotubes by Iijima in 1991 [1], there has been much research on this low-dimensional system because of its unusual physical properties [2-6] and its potential industrial applications [7]. Various methods have now been used to produce both single-wall and multi-wall nanotubes [8], and high quality aligned nanotube bundles [9]. Early on it was shown theoretically and experimentally that nanotubes with non-zero helicity are predominantly semiconducting whereas nanotubes of zero helicity are metallic [10, 11].

One aspect of carbon nanotube research exploits its hollowness to produce nanotubes filled with a variety of metallic and non-metallic materials along the nanotube axis. Initial filling of the nanotubes by Pb and Bi was achieved by capillary action [12]. Subsequently filling of nanotubes with other foreign materials was greatly aided by two important discoveries, respectively, by Tsang *et al*. [13] who showed that the nanotube tips can be opened with treatment of nitric acid and by Guerret-Piecourt *et al*. [14] who showed that filled carbon nanotubes can be synthesized by adding the filling material to the electrodes of the carbon arc discharge tube used for producing the nanotube [1]. Using these and other related techniques carbon nanotubes filled with Ag [15], $C_{60}$ [16], Ni [17], Co [18], TiC [19], Se, S, Sb, and Ge [20], MnC [21], LaC [22] and other materials have been produced. The electronic and other properties of the filled nanotubes are now being studied extensively. Electrical and thermal properties of $C_{60}$ filled single-wall carbon nanotubes or carbon peapods as they are called, have been investigated by Vavro *et al*. [23]. The magnetic and hysteritic properties of Fe-filled nanotubes have been examined by Prados *et al*. [24]. Catalytic properties and their possible applications in electrochemical energy storage of some filled carbon nanotubes have been studied by Che *et al*. [25]. The electronic structure and optical properties of filled nanotubes have been investigated theoretically by Ostling *et al*. [26] and Garcia-Vidal *et al*. [27], respectively. Yoon *et al*. [28] have studied the energetics and packing of fullerenes in nanotube peapods.

In a previous publication [29] we have studied theoretically the effect of electron-phonon interaction on the G-band Raman spectra of metallic carbon nanotubes. We have shown that excitation of a low frequency optical plasmon in a metallic carbon nanotube can split the G-band Raman line into two via the electron phonon interaction. This process successfully explained the observation of the Breit-Wigner-Fano (BWF) line shape near the G-band Raman spectra of a metallic carbon nanotube [30-33]. An alternative explanation of the BWF peak was originally provided by Kempa [34] on the basis of excitation of a pseudo-acoustic plasmon in the presence of momentum dependent defects in the sample.

We have also studied the Raman spectra of a nanotube filled with atoms or molecules and have found that the phonon of the filling atom gives rise to an additional peak [35]. The positions, widths and relative strengths of the three peaks were found to depend strongly on the phonon frequencies, the strengths and forms of the electron-phonon as well as phonon-phonon interaction.

Although our previous work considered only one nanotube phonon (G-band), a pristine carbon nanotube has three prominent Raman active phonon modes [32, 33, 36]. The one with lower frequency is known as the radial breathing mode (RBM) corresponding to the radial oscillation of the carbon atoms and the other two with higher frequency are called the G-bands related to the tangential motion of the carbon atoms on the surface of the nanotube. The higher G-band frequency corresponds to oscillation of the carbon atom parallel to the nanotube axis and the lower G-band frequency is associated with the azimuthal oscillation of the carbon atom. In the present article we study both the RBM and G-band Raman spectra of unfilled and filled semiconducting and metallic single-wall carbon nanotubes in the presence of both electron-phonon and phonon-phonon interactions. We find that these interactions and the individual phonon and plasmon frequencies have important effects on the positions, relative strengths and widths of the Raman lines of a filled or unfilled carbon nanotube. In Section II we introduce the model and the basic theory for studying the present problem. In Section III we present representative results of our calculation and a brief discussion. Section IV presents summary and conclusions of this work.

## II. THEORY

The Raman scattering process in a material can be understood at the microscopic level in the following manner: the incident photon first polarizes the medium by producing an electron-hole pair. Because of the presence of electron-phonon interaction, the electron or the hole of the pair can either emit or absorb a phonon before recombining to produce the scattered photon. The phonon thus produced or absorbed, in its turn, may be renormalized because of electron-phonon and phonon-phonon interactions. It can be shown [29, 34] that the Raman intensity is given by spectral density function (i.e., the imaginary part of the renormalized phonon propagator) for vanishingly small values of the wave vector $q$, since $q$ associated with the photon responsible for the Raman effect is vanishingly small.

Thus in order to calculate the intensity of the Raman spectra in a carbon nanotube we have to calculate the renormalized phonon propagator incorporating the following: presence of three Raman active phonons in the nanotube, one phonon corresponding to the vibration of the filling atoms, the interaction of the electrons of the nanotube with its phonons and the interaction of the nanotube phonons with the phonon of the filling atom, all of which can be modeled by the following Hamiltonian:

$$H = H_e + H_p + H_{e\text{-}p}. \tag{1}$$

Here $H_e$ is the Hamiltonian for the electrons of the nanotube and can be written as

$$H_e = \sum_k \varepsilon_k c_k^+ c_k + \frac{1}{2}\sum_{Q,k,k'} V_Q c_{k+Q}^+ c_{k'-Q}^+ c_{k'} c_k, \tag{2}$$

where $c_k(c_k^+)$ are the annihilation (creation) operator for the electrons on the nanotube. The first term on the right hand side represents the kinetic energy of the electrons and the second term is the Coulomb interaction between the electrons. Here the subscript $k, Q$ and $k'$ stand for both the wave vectors and the spin of the electrons. As will be seen below, these wave vectors will have two components, one for motion along the axis of the nanotube and one for azimuthal (perpendicular to axis) motion. The Hamiltonian $H_p$ in Eq. (1) is for the phonons of the nanotube and that of the filling atoms and can be written as

$$H_p = \sum_{q,i} \omega_{qi} b_{qi}^+ b_{qi} + N\omega_0 b_0^+ b_0 + \sum_{q,i} K A_{qi}^+ A_0 \tag{3}$$

where $A_{qi} = (b_{qi} + b_{-qi}^+)$ is the $q$th Fourier component of the displacements of the carbon atoms of the nanotube, $b_{qi}$ ($b_{qi}^+$) being the annihilation (creation) operator of the $i$-th phonon with the wave vector $q$ and $b_0$ ($b_0^+$) are the annihilation (creation) operator for the localized phonons of the filling atoms vibrating with a fixed frequency $\omega_0$, $N$ is the number of filling atoms, and $\omega_{qi}$ is the $i$-th phonon frequency of the carbon nanotube. Note that in this case we have three values of $i$ (=1, 2 and 3), since we are considering one RBM and two G-band phonons of the nanotube. In Eq. (3) $K$ is the strength of the interaction between the localized phonon mode of the filling atoms and the propagating phonon modes of the nanotube. In Fig. 1 we have provided a schematic diagram of a filled nanotube where we have assumed that each filling atom shown along the axis of the nanotube is harmonically coupled to the adjacent carbon atoms of the nanotube and can therefore vibrate with a frequency $\omega_0$ as mentioned above. The coupling constant for the harmonic coupling between the filling atom and the nanotube carbon atom will obviously depend on the nature of the filling atoms. In reference 28 the coupling constant between the filling molecule of fullerene and the nanotube has been estimated to be 041N/m.

The Hamiltonian representing the interaction between the electrons and the phonons of the nanotubes in Eq. (1) can be expressed as

$$H_{e-p} = g \sum_{k,q,i} c_{k+q}^+ c_k A_{qi}, \tag{4}$$

where g is the strength of the electron-phonon interaction.

Using the equation of motion method with the above Hamiltonians, the renormalized phonon Green's function $D_Q(\omega)$ for the nanotube has been calculated taking into account of the phonon-phonon interaction and the electron-phonon interaction within the random phase approximation and is given by

$$D_Q(\omega) = \frac{1}{[D_q^0(\omega)]^{-1} - (2\pi K)^2 D(\omega) - (2\pi g)^2 \chi_Q(\omega)} \, , \qquad (5)$$

where the wave vector $Q=[q,\mu/a]$ is associated with the motion of the electrons on the nanotube. Here $q$ is the component of the wave vector in the direction of the nanotube axis and $\mu$ is the quantum number associated with the azimuthal motion of the electrons on the surface of the nanotube. The first term in the denominator of Eq. (5) is the inverse of the bare phonon Green's function $D_q^0(\omega)$ for the three nanotube phonons and is given by

$$D_q^0(\omega) = \sum_{i=1}^{3} \frac{\omega_{qi}}{\pi(\omega^2 - \omega_{qi}^2)} \, . \qquad (6)$$

The second term in the denominator of Eq. (5) represents the effect of the interaction of the nanotube phonon with that of the filled atom, where the phonon Green's function for the filling atoms is approximately given by

$$D(\omega) \approx \frac{\omega_0}{\pi(\omega^2 - \omega_0^2)} \, , \qquad (7)$$

The third term in the denominator of Eq. (5) is the contribution of the electron-phonon interaction. It has been expressed in terms of the electric susceptibility $\chi_Q(\omega)$ due to electron polarization of the nanotube, which when calculated in the random phase approximation is

$$\chi_Q(\omega) = \frac{\Pi_Q(\omega)}{1 + V(Q)\Pi_Q(\omega)} \, , \qquad (8)$$

where $\Pi_Q(\omega)$ is the electron polarization propagator which for the metallic nanotubes has been previously shown to be [37]

$$\operatorname{Re}\Pi_Q(\omega) \approx -\frac{k_F^2}{2\pi m \omega^2}\left(q^2 + \frac{\mu^2}{a^2}\right) \, . \qquad (9)$$

Here $m$ and $a$ are the electron mass and radius of the nanotube, respectively and $k_F$ is the Fermi momentum of the nanotube electrons. Note that the imaginary part of $\Pi_Q(\omega)$ in the range of plasmon excitation is vanishingly small and will be given an infinitesimal value of $\varepsilon$. $V(Q)$ in Eq. (8) is the bare Coulomb interaction which for the electrons on the surface of a nanotube is known to be [37]

$$V(Q) = 4\pi e^2 a I_\mu(aq) K_\mu(aq), \qquad (10)$$

where $I_\mu$ and $K_\mu$ are the cylindrical Bessel functions and $e$ is the electronic charge. It should be mentioned that if the cylindrical geometry was not included in the calculation, the polarization propagator $\Pi_Q(\omega)$ and hence $\chi_Q(\omega)$ would be proportional to $q^2$ which would tend to zero for small $q$ and the Raman line would not get shifted by the electron-phonon interaction and there will be no BWF line shape due to plasmon excitation. This is exactly what happens in a three-dimensional normal metal. For the cylindrical geometry, however, Re $\Pi_Q(\omega)$ and Re $\chi_Q(\omega)$ are nonzero even when $q=0$ as can be seen from Eqs. (9) and (8) and hence the Raman lines would be shifted and there will be a BWF peak because of the electron-phonon interaction as shown below.

Substituting Eqs. (9) and (10) into Eq. (8) and taking the limit $q \to 0$, we find that the dimensionless electronic susceptibility of the nanotube is

$$\tilde{\chi}_Q(\omega)\Big|_{q=0} \equiv \tilde{\chi}_Q(\omega)/N(0)\Big|_{q=0} = \frac{\omega_p^2 \mu^2 a_0/a - i\tilde{\varepsilon}\omega^2}{(\omega^2 - \omega_p^2 + i\dfrac{a}{a_0}\tilde{\varepsilon}\omega^2)}, \qquad (11)$$

where $N(0) = \dfrac{m}{2\pi}$ is the density of states at the Fermi surface of the two-dimensional system, $a_0$ is the Bohr radius, $\tilde{\varepsilon} = \varepsilon/N(0)$ is the dimensionless infinitesimal imaginary part of the polarization propagator and $\omega_p^2 = \dfrac{\pi n e^2}{ma}$ is associated with the classical plasma frequency in a two-dimensional system. Here $n$ is the two-dimensional number density of the electrons in the nanotube. In Eq. (11) and below the tilde on top of any symbol would indicate that the quantity has been expressed in dimensionless units.

As we have explained above, the Raman scattering intensity is related to the spectral density function (or the imaginary part) of the phonon propagator $D_Q(\omega)$ of the nanotube in the limit of $q \to 0$ and hence can be written as

$$I(\omega) \propto \mathrm{Im}\, D_Q(\omega)\big|_{q=0}, \tag{12}$$

where $D_Q(\omega)$ is given by Eq. (5). Substituting Eqs. (6), (7), and (8) in Eq. (5), we obtain the Raman intensity from Eq. (12) as

$$\begin{aligned} I(\omega) &\propto -\mathrm{Im}\, \pi \omega_{q1} D_{q=0}(\omega + i\eta) \\ &= \mathrm{Im}\, \frac{1}{\{[\dfrac{\omega_{q1}}{(\omega+i\eta_1)^2 - \omega_{q1}^2} + \dfrac{\omega_{q2}}{(\omega+i\eta_2)^2 - \omega_{q2}^2} + \dfrac{\omega_{q3}}{(\omega+i\eta_3)^2 - \omega_{q3}^2}]^{-1} - \dfrac{4K^2 \omega_0}{(\omega+i\eta_0)^2 - \omega_0^2} - \dfrac{\tilde{\lambda}(\omega_p^2 \mu^2 a_0 / a - i\tilde{\varepsilon}\omega^2)}{\omega^2 - \omega_p^2 + i\tilde{\varepsilon}\omega^2 a / a_0}\}} \end{aligned} \tag{13}$$

where $\omega_{q1}$, $\omega_{q2}$ and $\omega_{q3}$ are the frequencies of the two G-band modes and the RBM mode of the nanotube, respectively, $\omega_0$ is the frequency associated with phonon of the filling atom and $\tilde{\lambda} = (\dfrac{\pi N(0) g^2}{\omega_{q1}})$ is the dimensionless parameter determining the strength of the electron-phonon interaction. Note that here a parameter $\eta$ has been introduced as the natural infinitesimal width of all the bare phonon propagators by replacing every $\omega$ in the propagators by $\omega + i\eta$. Furthermore, in this paper we will normalize all frequency parameters with respect to the frequency for the higher G-mode $\omega_{q1}$ and as mentioned before, we represent the dimensionless quantities thus created by a tilde.

A careful examination of Eq. (13) will reveal that the Raman intensity will, in general, have five peaks near the normalized frequencies of $\tilde{\omega}_{q1}$ (=1), $\tilde{\omega}_{q2}$, $\tilde{\omega}_{q3}$ $\mu \tilde{\omega}_p$ and $\tilde{\omega}_0$, corresponding to the phonon excitations of the nanotube (two G-band and one RBM phonons), the plasmon, and the localized mode of the filling atom, respectively. However, some of these peaks may get suppressed and others may get enhanced because of strong electron-phonon and phonon-phonon interactions. Note that for $\mu=0$, the peak near $\mu\tilde{\omega}_p$ will drop out, which indicates that the pseudo-acoustic plasmon mode [29] corresponding to $\mu=0$ will not modify the G-bands to produce the BWF peak. We have therefore worked with non-zero values of $\mu$ which will correspond to excitation of an optical plasmon. For $\tilde{\omega}_0 = 0$, the peak due to the localized phonon mode of the filling atom will not be present which will correspond to the case of an unfilled nanotube.

## III. RESULTS AND DISCUSSION

The Raman intensity as given by Eq. (13) will depend on the parameters $\tilde{\lambda}$, $a_0/a$, $\mu$, $\tilde{\omega}_{q1}$, $\tilde{\omega}_{q2}$, $\tilde{\omega}_{q3}$, $\tilde{\omega}_p$, $\tilde{K}^2$, $\tilde{\omega}_0$, $\tilde{\eta}_1$, $\tilde{\eta}_2$, $\tilde{\eta}_3$ and $\tilde{\varepsilon}$. Higher G-band phonon frequency $\tilde{\omega}_{q1}$ is known to be independent of nanotube radius whereas both the second G band frequency $\tilde{\omega}_{q2}$ corresponding to the azimuthal motion of the carbon atom and the RBM frequency $\tilde{\omega}_{q3}$ are radius dependent [33, 38]. In this paper the normalized dimensionless higher G-band phonon frequency $\tilde{\omega}_{q1}$ corresponding to *1591* cm$^{-1}$ is taken to be *1*. Then following reference [33], we find that the dimensionless second G-band frequency is given by $\tilde{\omega}_{q2} = 1 - 2.6683(\frac{a_0}{a})^2$. The dimensionless RBM frequencies $\tilde{\omega}_{q3}$ corresponding to different radii of the nanotube are shown in the relevant figures. We have plotted the Raman intensity as a function of the frequency for several representative sets of these parameters. In all of our plots we have taken $\mu=1$ as we expect that, other than $\mu=0$ which neither produces the BWF peak nor shifts the Raman lines, this being the lowest energy optic like mode will be the easiest plasmon mode to excite.

In Figs. 2a and b, we have plotted the Raman spectra of an unfilled ($\tilde{K}^2=0$, $\tilde{\omega}_0=0$) semiconducting nanotube with three different values of the nanotube radius and hence three values of the lower G-band and three reasonably consistent values of the RBM frequency. Values of all the relevant parameters are shown in the legends of this and all other figures, although we have omitted the tilde symbols in the figures for simplicity of notation. Note that in the semiconducting sample there will be no plasmon excitation ($\tilde{\omega}_p = 0$ and $\tilde{\lambda} = 0$) and therefore, the two G-bands appearing near $\tilde{\omega} = 1$ have Lorentzian form, and lower G band appears at different frequencies because of their radius dependence as clearly seen in Fig. 2b. This result has been verified by experiments [32]. The lower frequency RBM band appearing near $\tilde{\omega} = 0.1$ also has a Lorentzian form and appears at three different frequencies because of their radius dependence [33, 38]. Please note that in Fig. 2 and all the subsequent figures the Raman intensity (I) has been plotted in arbitrary units as a function of the normalized frequency (ω) expressed in dimensionless units.

In Figs. 3a and b we have plotted the Raman spectra for an unfilled metallic nanotube for three values of the nanotube radius and hence three values of lower G-band frequency and three values of RBM frequency. Note that the presence of the electron phonon interaction (nonzero $\tilde{\lambda}$ and $\tilde{\omega}_p$) drastically modifies the G-band spectra. As seen in Fig. 3b the lower G band appearing near $\tilde{\omega} = 0.98$ becomes asymmetric and resembles the BWF line shape observed in experiments [30-33]. This asymmetric BWF line shape of the G-band in the metallic nanotube arises due to the proximity of the optical plasmon frequency ($\mu = 1$) to that of the lower G-band phonon. The upper G band, however, basically retains its Lorentzian shape. We also notice that the lower G band peak moves to higher frequency with the increase of the radius of the nanotube and such a shift has been observed in experiments [32]. Our theory indicates that the upper G band peak also moves with radius although this movement is less pronounced. Furthermore, the electron-phonon interaction drastically reduces the G-band intensity whereas the low frequency RBM mode remains relatively unaltered.

We then considered the Raman spectra of a filled carbon nanotube. In Figs. 4a, b, and c we have plotted the Raman spectra for a filled metallic carbon nanotube for three different values of the frequency of vibration of the filling atom. This filling atom frequency has been chosen to be close to the RBM frequency - one is slightly lower, and the other two are slightly higher than the RBM frequency ($\tilde{\omega}_{q3} = 0.09$) as shown in the legend of Fig. 4a. All the other parameters are the same as those of Fig. 3. Although the G-bands remain practically unchanged, there is a big change in the RBM spectra. As can be clearly seen in Fig. 4b, where we have plotted the Raman intensity close to RBM frequencies, that phonon of the filling atom gives rise to an extra peak near the RBM peak. While the two G-bands are not perceptibly affected by the presence of the filling atom (Fig.3c), there is a strong interaction between the RBM mode and the phonon of the filling atom (Fig. 4b). The two peaks get hybridized and repel each other, the one associated with the filling atom moves to higher frequency whereas the one belonging to the RBM mode moves to lower frequency. When $\tilde{\omega}_0 \leq \tilde{\omega}_{q3}$, almost all the intensity gets transferred to the filling atom mode and the RBM intensity becomes flat and may become unobservable. But with $\tilde{\omega}_0$ increasing in value the two peaks move apart and the RBM mode gains intensity compared to the peak intensity of the filling atom mode. The behavior of the Raman spectra of a filled semiconducting nanotube is almost the same as that shown in Fig. 4 except that in this case, as expected, the BWF satellite will be missing and the two G-bands will be of Lorentzian form.

We have then investigated the effect of variation of the strength of the interaction $\tilde{K}^2$ between the nanotube phonons and the phonon of the filling atom on the Raman spectra of the filled nanotube. In Fig. 5 we have plotted the Raman spectra of a metallic nanotube for three different values of $\tilde{K}^2$ and for $\tilde{\omega}_0 = 0.16$, keeping all the other parameters the same as in Fig. 4. As in Fig. 4, since $\tilde{\omega}_0$ and $\tilde{\omega}_{q3}$ are much lower than the G-band frequencies, the two G-bands are not perceptibly affected by this variation and therefore we have not plotted them separately here. However, variation of $\tilde{K}^2$ seems to have a significant effect on the Raman spectra associated with the RBM and filling atom modes as shown more explicitly in Fig. 5b. The filling atom peak occurs at a higher frequency in all three cases. However, with decreasing $\tilde{K}^2$ the RBM mode gains strength compared to the filling atom mode. In the limit of $\tilde{K}^2$ approaching zero, the filling atom intensity will vanish and only the RBM band will survive. Calculation of the variation of the Raman spectra of a semiconducting filled nanotube with variation of $\tilde{K}^2$ shows similar behavior, and we have therefore not plotted them here.

Since many different kinds of elements have been used as the filling atoms [12-23], the phonon frequency of the filling atom can be quite different from that of the RBM. In such cases the filling atom modifies the Raman spectra of the nanotube differently. As an example, in Fig. 6 we have plotted the Raman spectra of a filled metallic nanotube for three values of the normalized filling atom frequencies $\tilde{\omega}_0$ (*0.4*, *0.6* and *0.8*) lying between the RBM and the G-band frequencies and for $\tilde{K}^2 =0.008$, with other parameters remaining the same as those of Fig. 5. As expected the filling atom gives rise to an extra peak at a frequency lying between RBM and G-band frequencies and also shifts and modifies the RBM and the two G-bands as can be seen Figs. 5b and 5c. Note that the higher frequency filling atom has a stronger effect on the G-bands moving them to higher frequencies as can be seen explicitly in Fig. 6b, whereas the lower frequency filling atom has a stronger effect on the RBM mode moving it to lower frequency as seen in Fig. 6c. The presence of filling atoms produces similar effects on the Raman spectra of semiconducting nanotubes, which have not been plotted here.

In the discussion above, it has been pointed out that in the case of filled semiconducting nanotubes the behavior of the Raman peaks corresponding to the RBM, the G-bands and the filled atom modes are similar to that of the metallic case, with the difference that both G-bands in the semiconducting nanotube have Lorentzian shape. While this conclusion is true in general, the situation can be quite different if the semiconducting nanotube is filled with donor atoms. In such a case, the conduction band of the semiconducting nanotube will be partially filled by donor electrons and the system will behave like a metal. There will then exist an optical plasmon mode ($\mu =1$) whose frequency will depend on the amount of filling (the filling fraction) of the nanotube. Interaction of this plasmon with the nanotube phonon will give rise to an extra peak in the Raman spectra of a semiconducting nanotube filled with donor atoms. If, for example, the filling fraction is such that the corresponding plasmon frequency is close to the RBM, or the G-band or the filling atom phonon frequency, it will give rise to a BWF like satellite close to the RBM band or the G-band, or the filling atom phonon peak, respectively. Thus in the semiconducting nanotubes filled with donor atoms, there is the possibility of observing the BWF like line shapes near the RBM, G-band or the filled atom modes depending upon the filling fraction.

Two of these possibilities are depicted in Figs. 6 and 7. In Fig. 7 we have plotted Raman spectra of a semiconducting nanotube filled with donor atoms for three different plasmon frequencies of the donor electrons close to the RBM frequency. Each plasmon frequency gives rise to an extra asymmetric BWF type peak close to the RBM peak as can be clearly seen in Fig. 7. In this case the G-bands and the peak corresponding to the filling atom phonon remain practically unaffected and so we have not plotted them here. In Fig. 8 we have plotted the Raman spectra of a filled semiconducting nanotube for three different plasmon frequencies of the donor electrons close to the phonon frequency of the filling atoms ($\tilde{\omega}_0 = 0.6$). In this case also we find three extra asymmetric peak close to the large intensity Raman peak of the filling atom, which can be clearly seen in this figure. In this case also the G-bands and the RBM peaks remain practically unaffected and so we have not plotted them here.

These predictions can be verified experimentally if one performs experiments on a semiconducting nanotube filled with donor atoms. It should be pointed out that this effect provides a unique method of determining the filling fraction of semiconducting carbon nanotubes and determining the electrical nature of the filling atoms.

## IV. SUMMARY AND CONCLUSIONS

In this paper we have studied theoretically the Raman spectra of unfilled and filled semiconducting and metallic carbon nanotubes in the presence of the electron-phonon and phonon-phonon interactions. For the unfilled semiconducting and metallic nanotube we have considered one lower frequency RBM mode and two prominent higher frequency G-band modes. As shown in Figs. 2 and 3, in the semiconducting nanotube all three Raman peaks are of the Lorentzian form whereas for the unfilled metallic nanotube the lower G-band is modified by the excitation of an optical plasmon making it asymmetric and wider. This asymmetric band has been designated as the BWF line and has previously been studied extensively both experimentally and theoretically.

Assuming that the filling atom gives rise to a localized mode which is a Raman active phonon, we find that the Raman spectra of a filled nanotube, in general, have an additional peak corresponding to the phonon of the filling atom. Depending on the frequency of this additional phonon and its strength of interaction with the nanotube phonons, we find some very interesting effects. For example, for a given strength of this interaction, if this phonon frequency is close to the RBM phonon frequency, there is strong hybridization of the Raman peaks of the RBM and that of the filling atom phonon. As the filling atom phonon frequency decreases and moves from above the RBM phonon frequency to below this frequency, the filling atom peak moves to lower frequency and gathers strength compared to the RBM peak as shown in Fig. 4. Eventually the filling atom mode becomes very prominent almost obliterating the RBM mode. For given values of the filling atom phonon frequency $\widetilde{\omega}_0$ and RBM frequency $\widetilde{\omega}_{q3}$ lying close to each other, when the phonon-phonon interaction strength ($\widetilde{K}^2$) is varied, the two peaks show sharp variation because of strong hybridization due to the phonon-phonon interaction as shown in Fig. 5. In the range of $\widetilde{K}^2$ shown in this figure both RBM peak (one on the left in Fig. 5b) and the filling atom peak (one on the right) move to the left as $\widetilde{K}^2$ decreases. For $\widetilde{K}^2 = 0$, the filling atom peak disappears and we are left with only the RBM peak only. In all these cases the high frequency G-bands line are not significantly affected by the presence of the low frequency filling atom mode. When the filling atom frequency is in the intermediate range, the effect of hybridization is less pronounced since the filling atom frequency is not close to either the RBM or the G-band frequency and hence the interaction is small (Figs. 5a). However, the effect of hybridization with G-bands is stronger when the filling atom frequency is closer to the G-band frequencies (Fig. 6b) and this effect is stronger with the RBM when the filling atom frequency is closer to the RBM frequency.

In semiconducting nanotubes filled with non-donor atoms, the behavior of the Raman peaks is similar to that of the metallic case, with the difference that the both G-bands are of Lorentzian form. However, it is predicted that if the filling atoms are donors, then because of tunability of the plasmon frequency with filling fraction, even in a semiconducting nanotube the BWF like line shape can appear close to the RBM (Fig. 7) or the G-band (similar to Fig. 3) or the filling atom mode Raman peaks (Fig. 8), depending upon the plasmon frequency of the donor electrons. This observation can become a valuable tool for determining the filling fraction and the electrical nature of the filling atoms in semiconducting carbon nanotubes.

Our results for unfilled metallic or semiconducting nanotubes are similar to those observed experimentally [30-33]. However, we are not aware of any experimental work on the Raman spectra of a filled nanotube. Our expectation is that carefully conducted experiments on the Raman spectra of filled nanotubes should show the pattern of results that we have discussed in this paper.

## ACKNOWLEDGMENT

One of us (SNB) would like to acknowledge the hospitality of the Physics Department of Drexel University where bulk of the work was carried out.

# FIGURE CAPTIONS

Fig. 1. Schematic representation of a filled single-wall carbon nanotube [39]. It is assumed that each filling atom is harmonically coupled to the carbon atoms of the nanotube.

Fig. 2. (Color online) (a) The RBM and two G-bands Raman spectra of an unfilled semiconducting nanotube for three values of nanotube radius and hence three values of lower G-band and RBM frequencies. (b) Raman spectra of unfilled semiconducting nanotube near the G-band frequencies. Note that all three bands are Lorentzian in form. In Fig. 2 and in all the subsequent figures, the Raman intensity (I) has been expressed in arbitrary units and the normalized frequency $\omega$ has been plotted in dimensionless units.

Fig. 3. (Color online) (a) The RBM and two G-bands Raman spectra of an unfilled metallic carbon nanotube for three values of the radius of the nanotube and the corresponding three RBM and three lower G-band frequencies. (b) The metallic nanotubes clearly show the asymmetric and wider BWF satellite band close to the upper G-band.

Fig. 4. (Color online) (a) Raman spectra of a filled metallic carbon nanotube with three values of the filling atom phonon frequencies close to the RBM frequency of *0.09*. There is an extra peak due to the filling atom phonon which hybridizes with the peak of the RBM phonon. (b) Raman spectra near the RBM mode. (c) Raman spectra near the G-bands. Note that the G-bands are not affected appreciably by the low frequency filling atom.

Fig. 5. (Color online) (a) Raman spectra of a filled metallic carbon nanotube for three different values of $K^2$, with the filling atom frequency close to the RBM frequency. (b) Raman spectra near the RBM mode. Notice strong effect of hybridization.

Fig. 6. (Color online) (a) Raman spectra of a filled metallic carbon nanotube with three values of the filling atom frequencies, lying between the RBM frequency and the G-band frequencies. (b) Raman spectra close to the G-bands. (c) Raman spectra close to the RBM band.

Fig. 7. (Color online) Raman spectra of a semiconducting nanotube filled with donor atoms. Three different donor electron plasmon frequencies have been chosen close to the RBM frequency. BWF like satellite peaks appear close to RBM Raman peak due to excitation of plasmons of the donor electrons

Fig. 8. (Color online) Raman spectra of a semiconducting nanotube filled with donor atoms. Three different donor electron plasmon frequencies have been chosen to be close to the filling atom phonon frequency ($\tilde{\omega}_0 = 0.6$). Notice the BWF like peaks near the filling atom phonon peak due to excitation of plasmons of donor electrons.

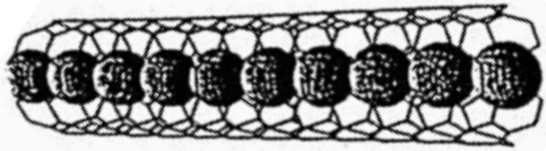

Figure 1

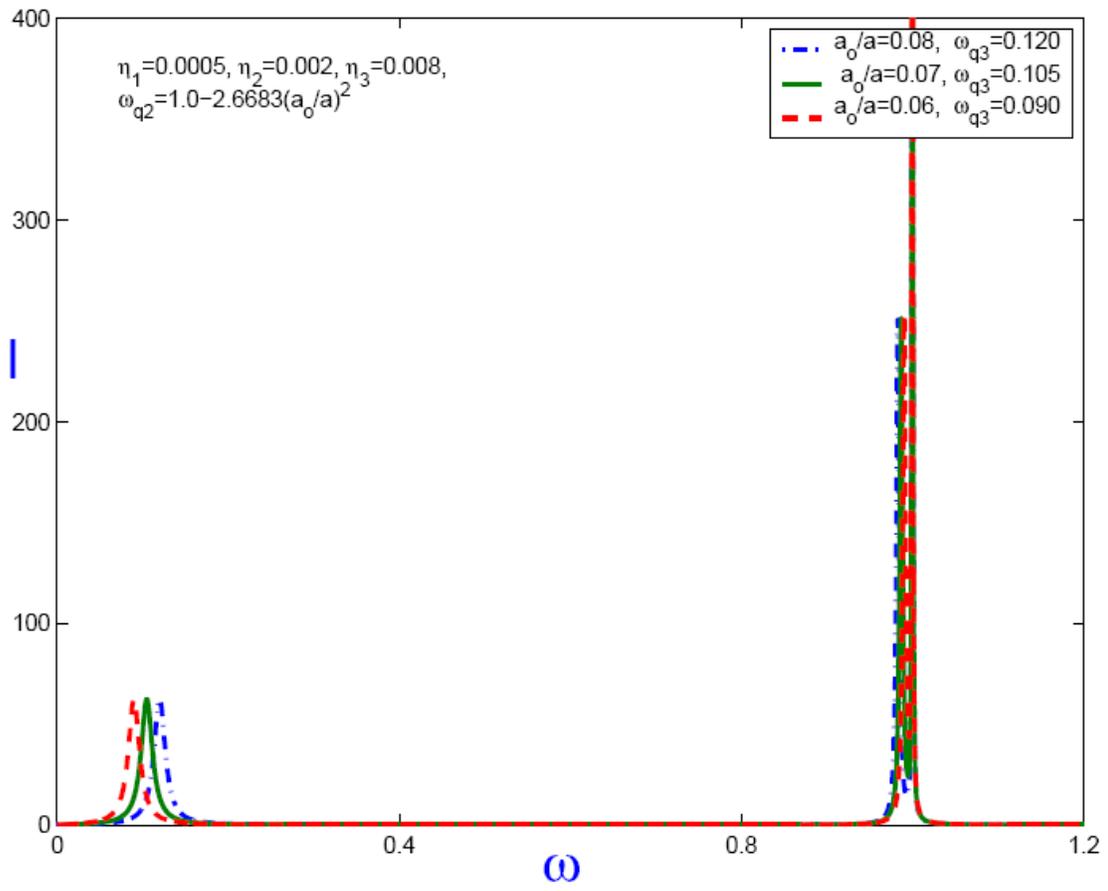

Figure 2a

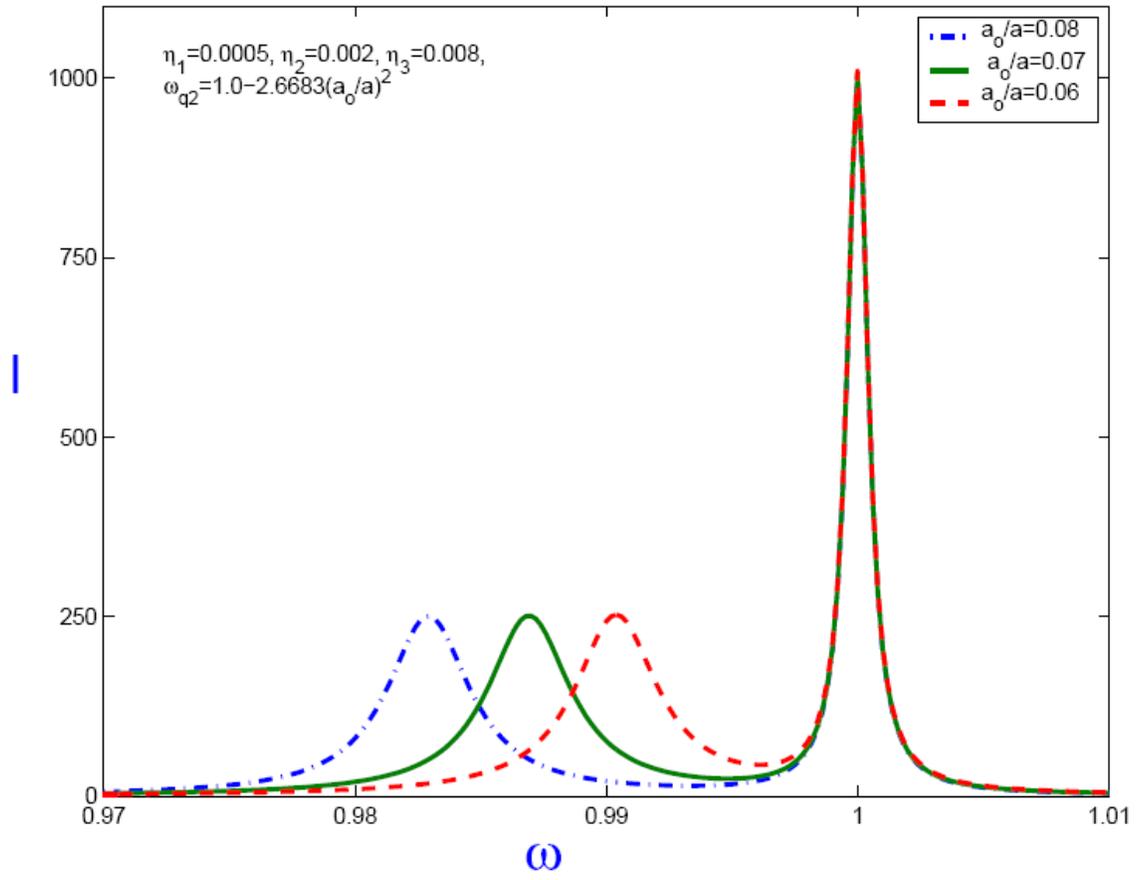

Figure 2b

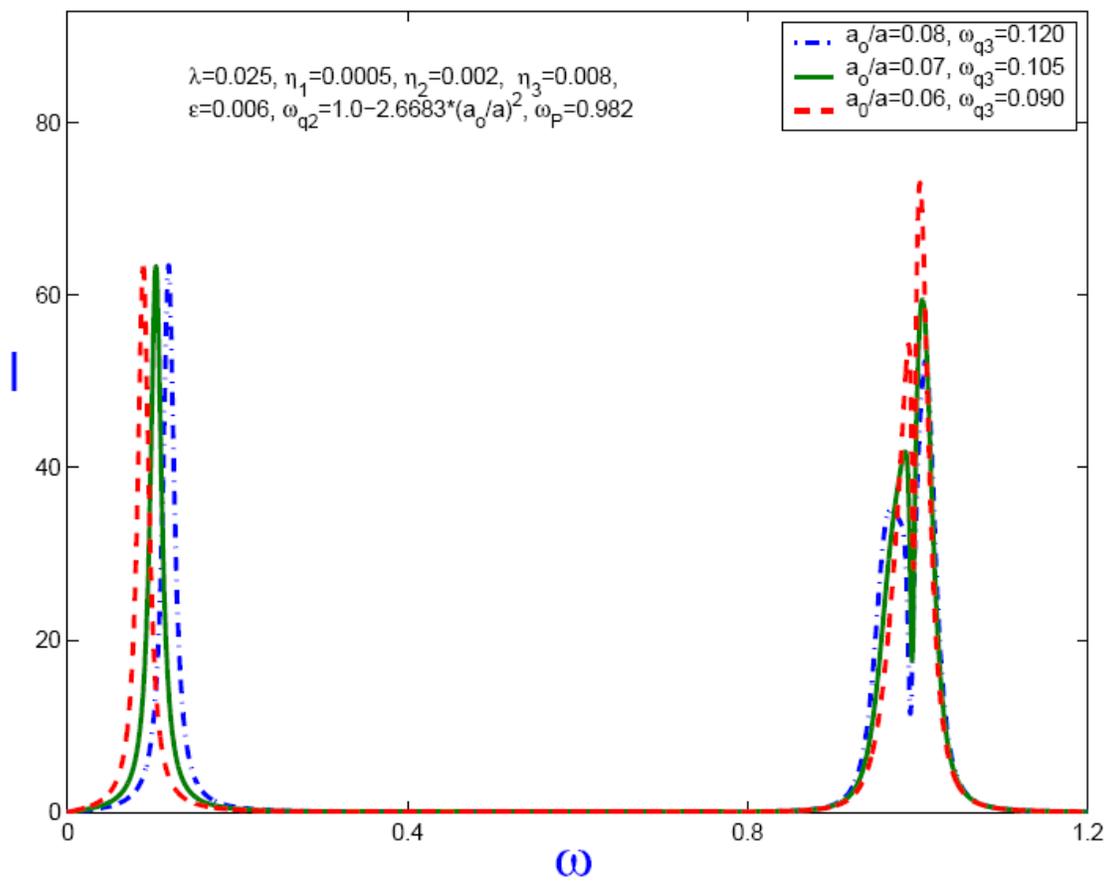

Figure 3a

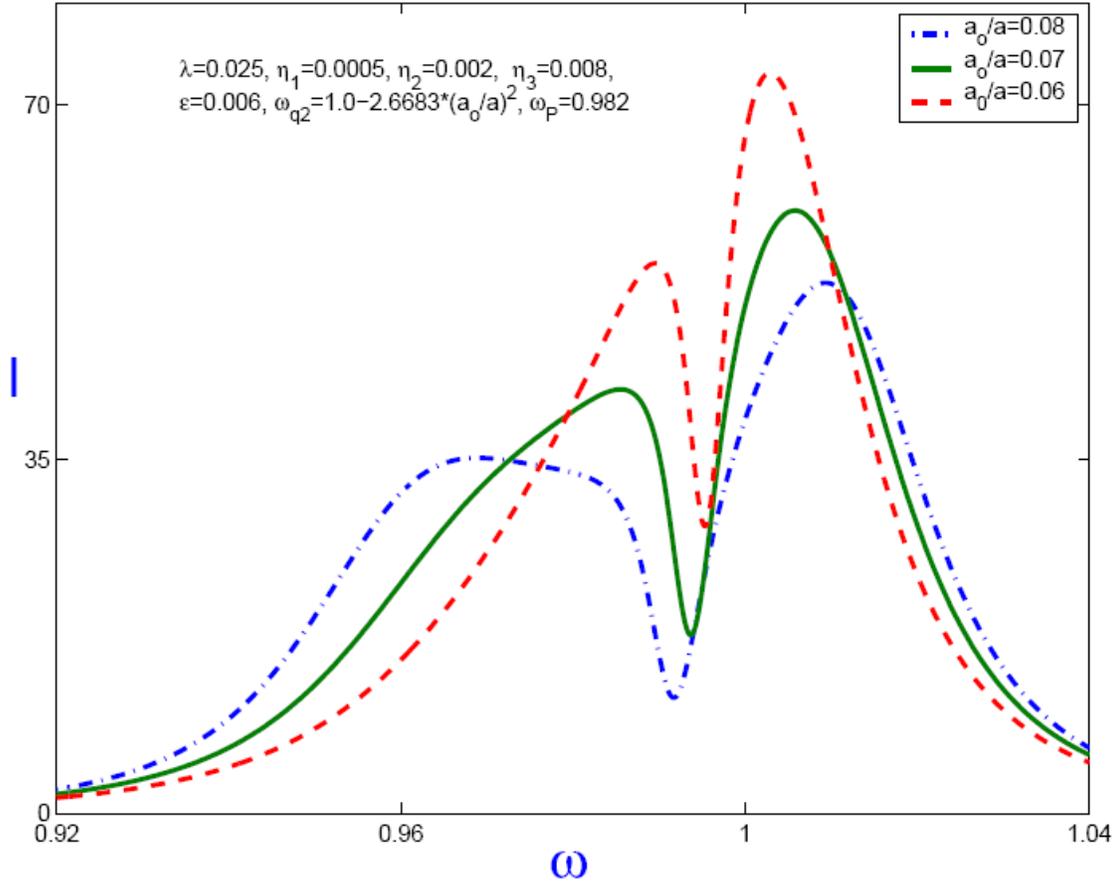

Figure 3b

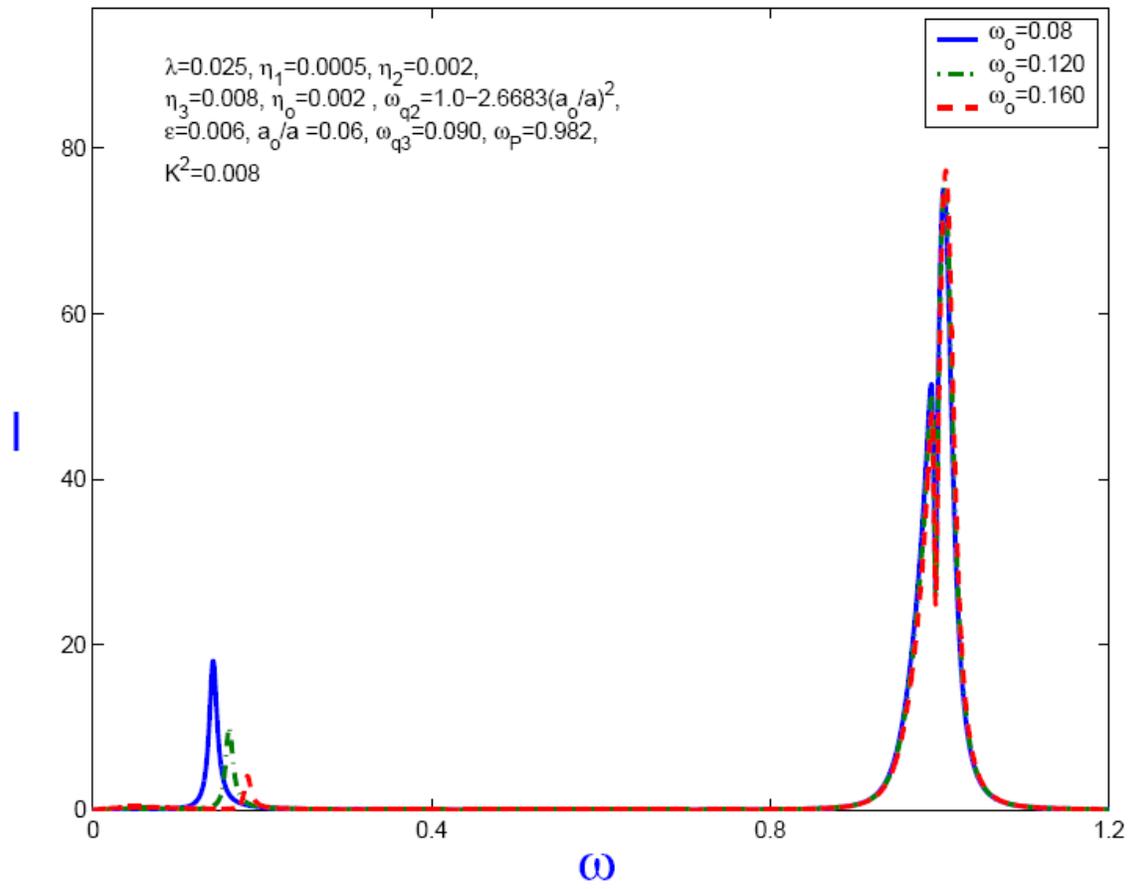

Figure 4a

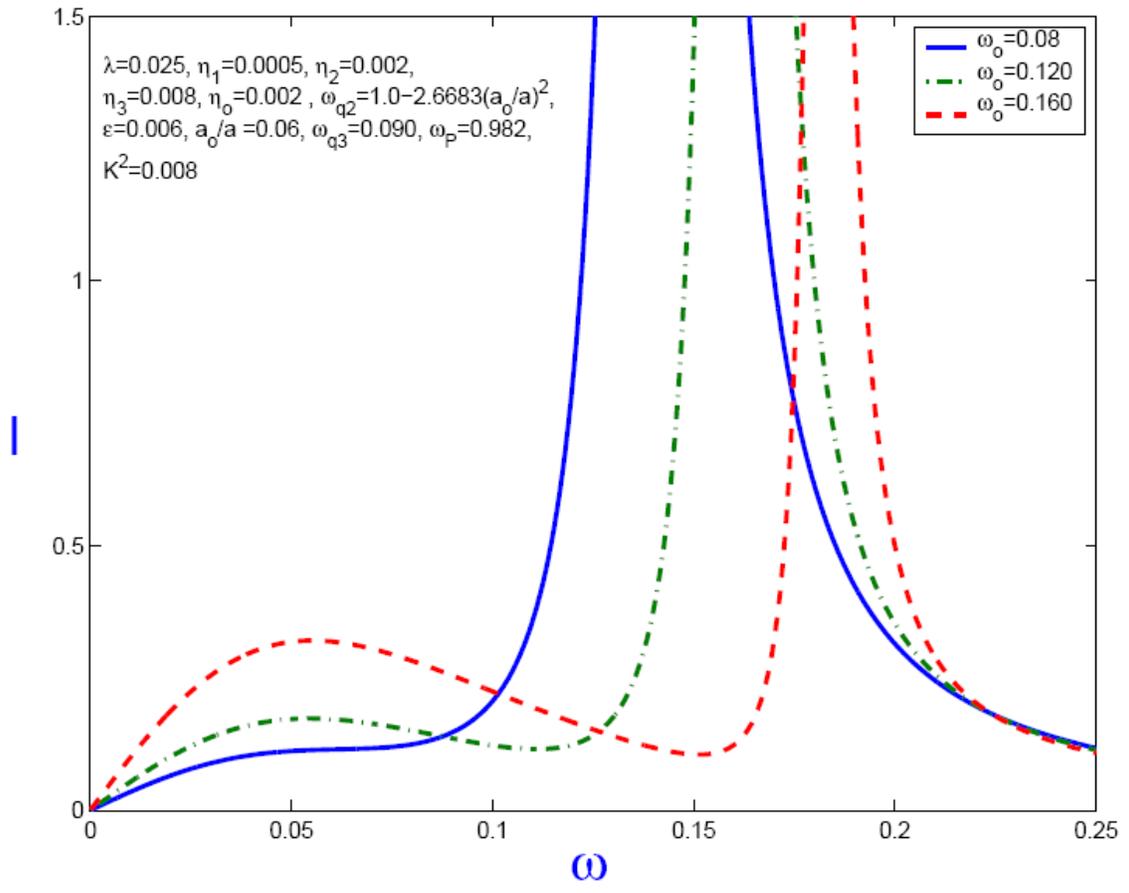

Figure 4b

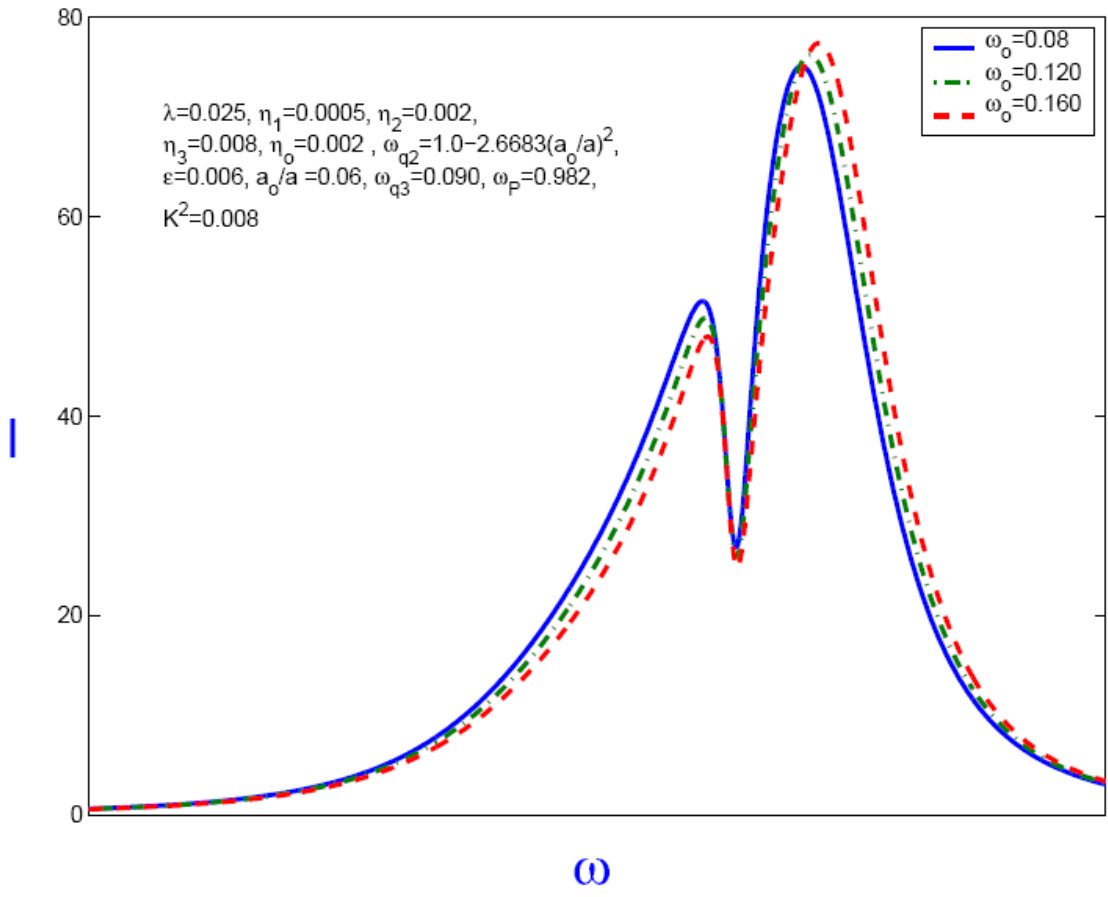

Figure 4c

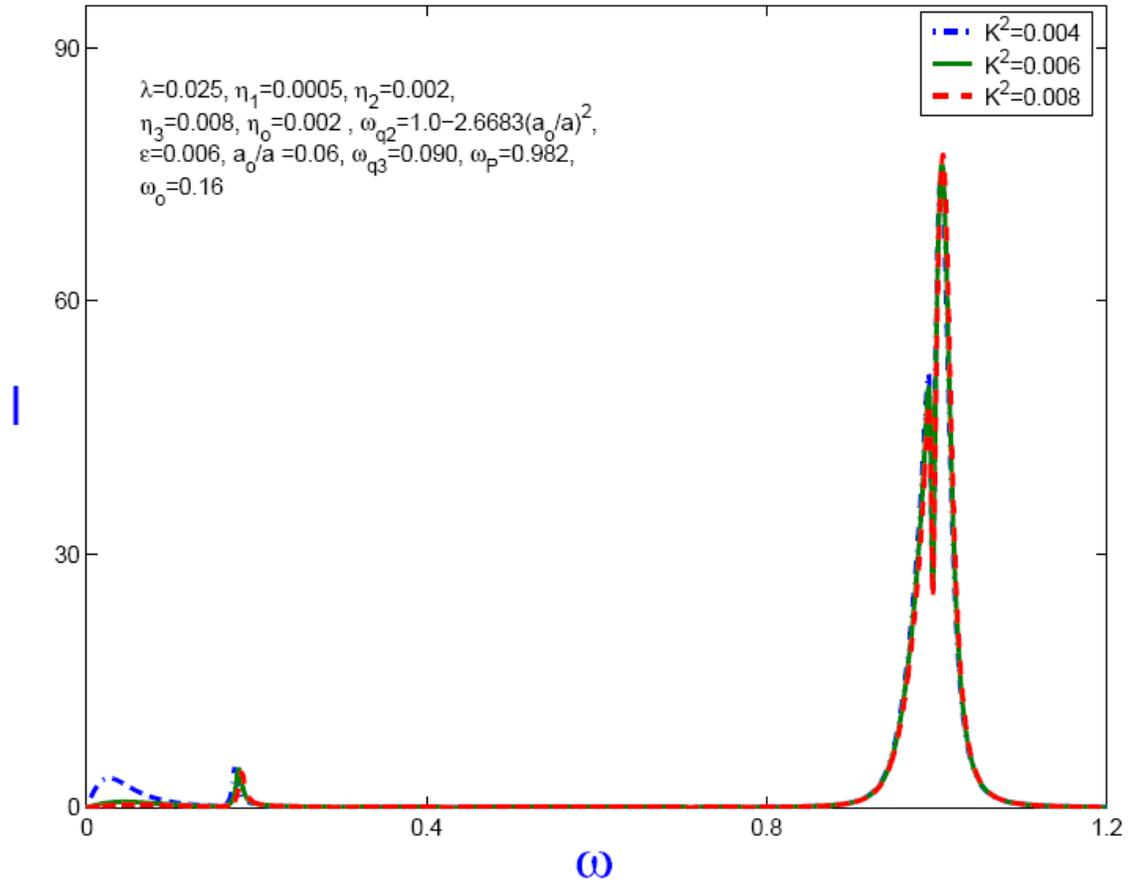

Figure 5a

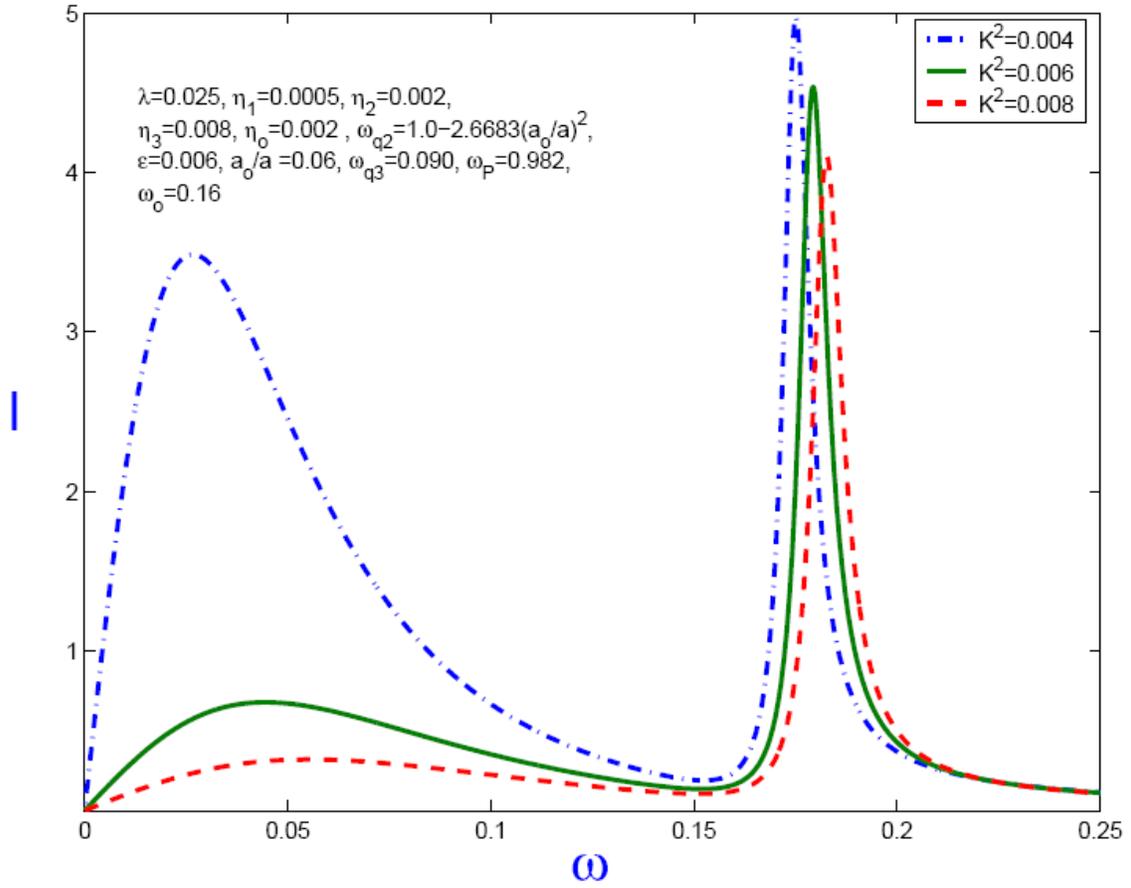

Figure 5b

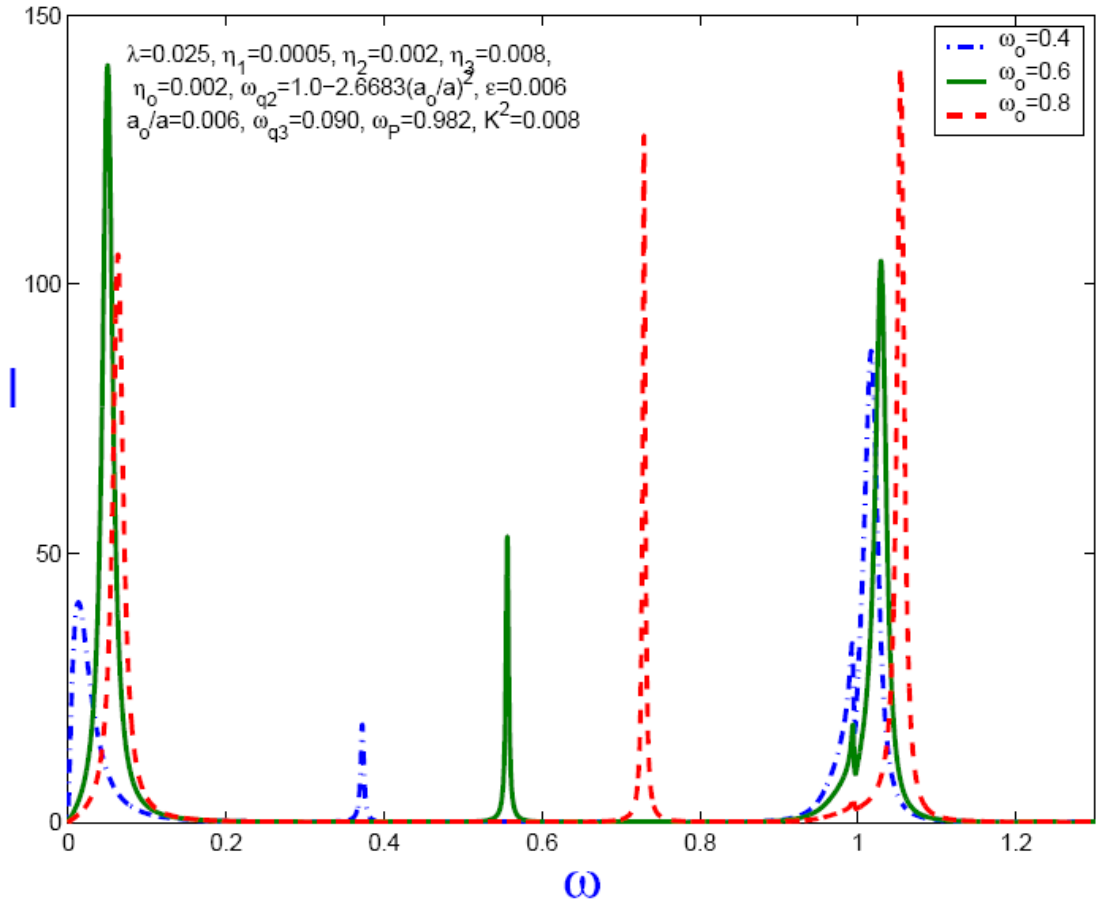

Figure 6a

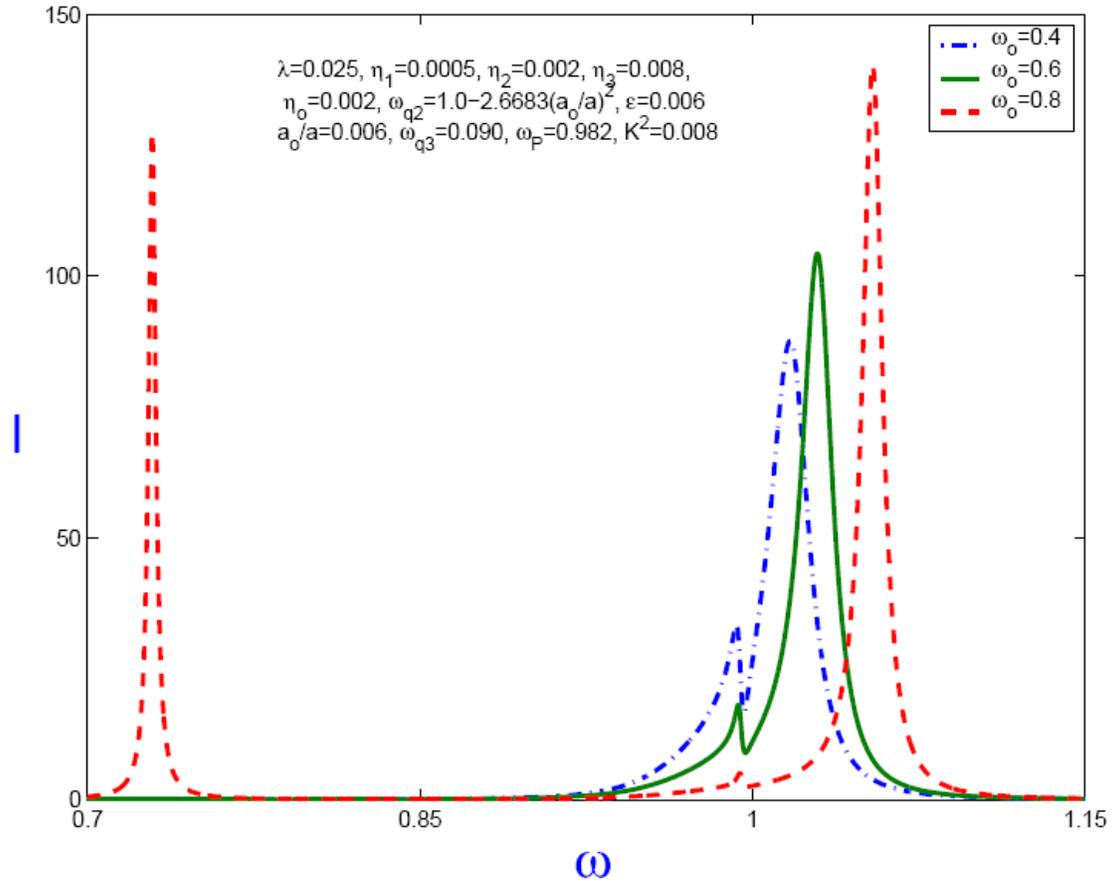

Figure 6b

<br />

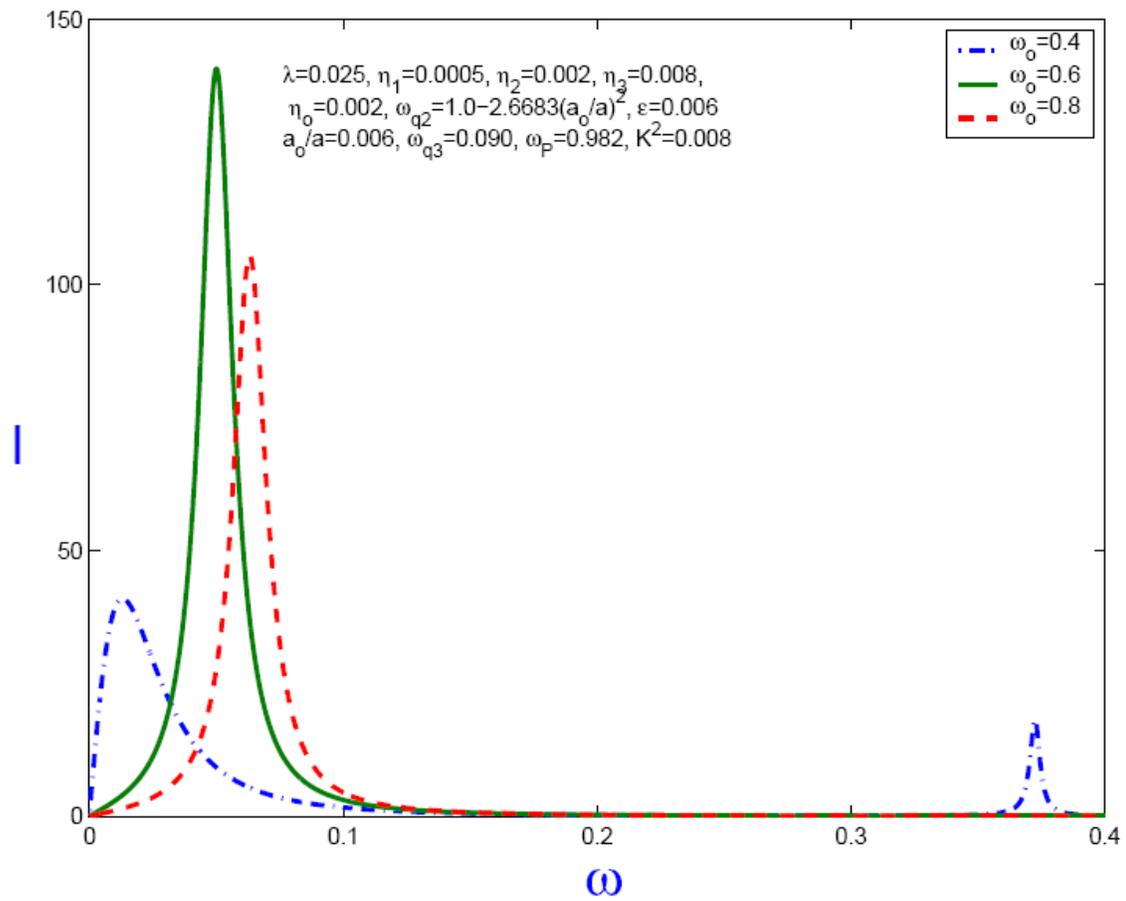

Figure 6c

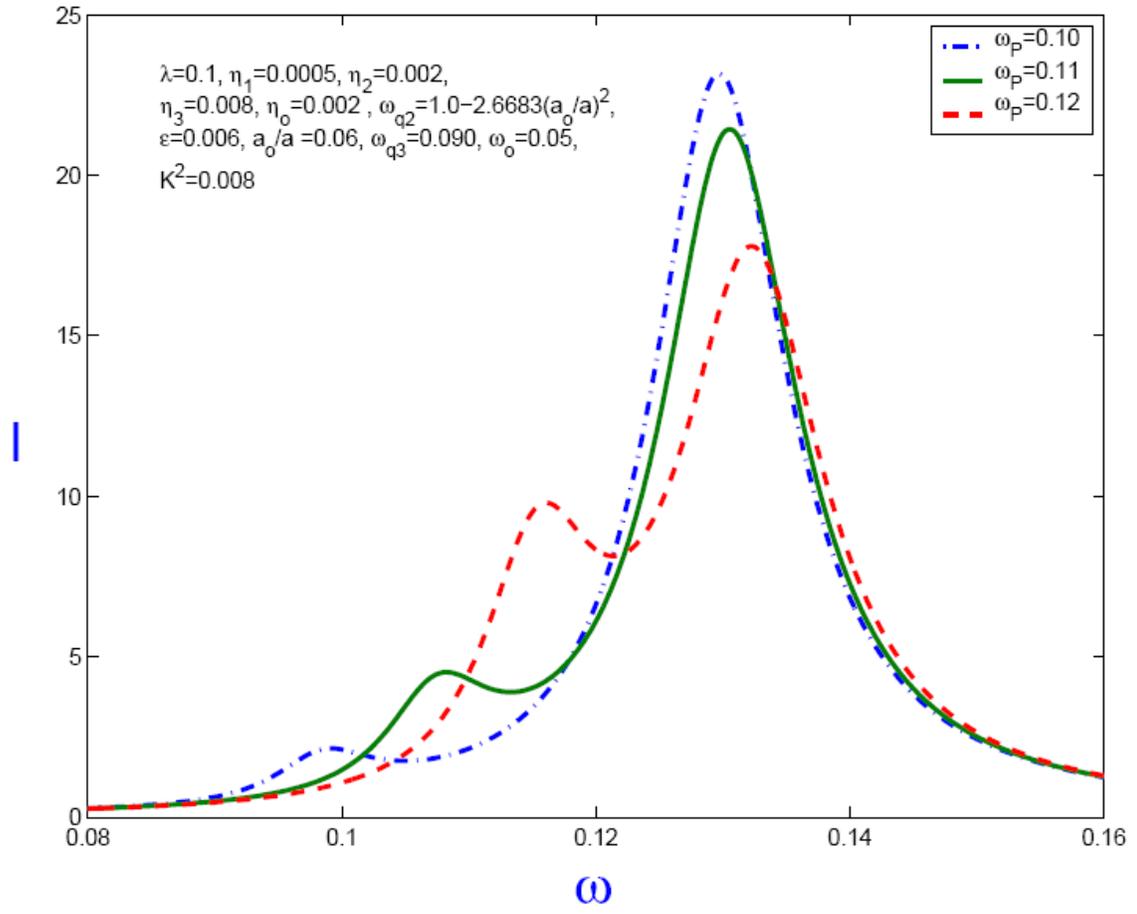

Figure 7

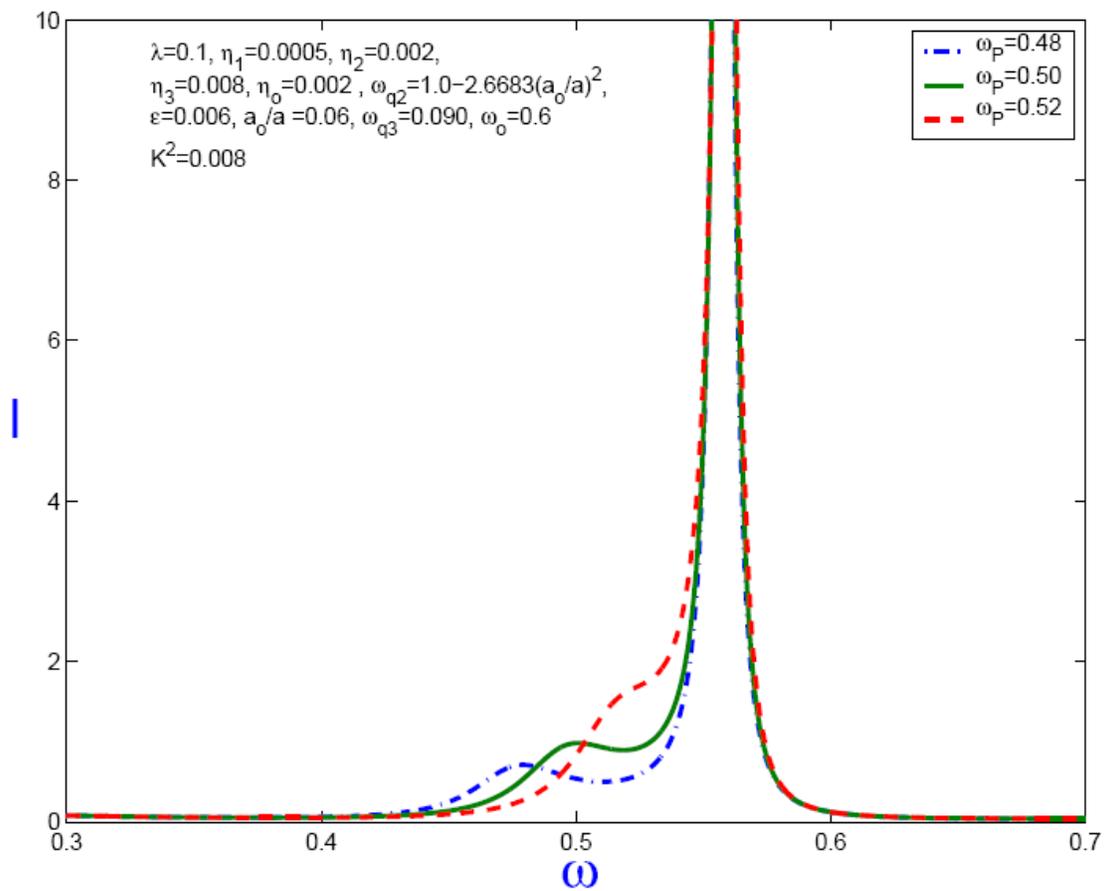

Figure 8